# Environment-insensitive and gate-controllable photocurrent enabled by bandgap engineering of MoS$_2$ junctions

Fu-Yu Shih[1,2], Yueh-Chun Wu[2,§], Yi-Siang Shih[1], Ming-Chiuan Shih[3], Tsuei-Shin Wu[1,2], Po-Hsun Ho[4], Chun-Wei Chen[4], Yang-Fang Chen[1], Ya-Ping Chiu[1,5], and Wei-Hua Wang[2*]

[1]Department of Physics, National Taiwan University, Taipei 106, Taiwan

[2]Institute of Atomic and Molecular Sciences, Academia Sinica, Taipei 106, Taiwan

[3]Department of Physics, National Sun Yat-sen University, Kaohsiung, Taiwan

[4]Department of Materials Science and Engineering, National Taiwan University, Taipei 106, Taiwan

[5]Institute of Physics, Academia Sinica, Taipei 115, Taiwan

[§]Current address: Department of Physics, University of Massachusetts, Amherst, Massachusetts 01003, United States

[*]Corresponding Author. (W.-H. Wang) Tel: +886-2-2366-8208, Fax: +886-2-2362-0200;

E-mail: wwang@sinica.edu.tw

Two-dimensional (2D) materials are composed of atomically thin crystals with an enormous surface-to-volume ratio, and their physical properties can be easily subjected to the change of the chemical environment. Encapsulation with other layered materials, such as hexagonal boron nitride, is a common practice; however, this approach often requires inextricable fabrication processes. Alternatively, it is intriguing to explore methods to control transport properties in the circumstance of no encapsulated layer. This is very challenging because of the ubiquitous presence of adsorbents, which can lead to charged-impurity scattering sites, charge traps, and recombination centers. Here, we show that the short-circuit photocurrent originated from the built-in electric field at the $MoS_2$ junction is surprisingly insensitive to the gaseous environment over the range from a vacuum of $1 \times 10^{-6}$ Torr to ambient condition. The environmental insensitivity of the short-circuit photocurrent is attributed to the characteristic of the diffusion current that is associated with the gradient of carrier density. Conversely, the photocurrent with bias exhibits typical persistent photoconductivity and greatly depends on the gaseous environment. The observation of environment-insensitive short-circuit photocurrent demonstrates an alternative method to design device structure for 2D-material-based optoelectronic applications.

**Introduction**

Generally, two-dimensional (2D) materials, composed of atomically thin crystals, exhibit an enormous surface-to-volume ratio, and the physical properties of 2D materials, including electrical, optical, and mechanical properties, are easily subjected to the change of the chemical environment.[1-6] Regarding the electrical property, the carrier transport in 2D materials is very sensitive to the presence of extrinsic adsorbents, which typically cause charged-impurity scattering, charge trapping, and recombination centers,[7-11] leading to degradation of the transport characteristics.[12-14] Although various encapsulation methods have been developed,[15-19] it is intriguing to explore methods to control the transport properties in the circumstance of no encapsulated layer. Here, we demonstrated that the short-circuit photocurrent enabled by the built-in electric field at the $MoS_2$ junction is surprisingly insensitive to the gaseous environment, which is very uncommon in the photoresponse of thin transition-metal dichalcogenides (TMDCs). We exploit the unique property of 2D TMDCs, in which the electronic band structures are associated with the number of the layered materials,[20,21] to create a junction structure. For molybdenum disulfide ($MoS_2$), the transition changes from direct bandgap (1.9 eV) in monolayer $MoS_2$ to indirect bandgap (1.3 eV) in bulk $MoS_2$.[22-24] This layer-dependent electronic structure therefore offers a distinct approach for designing the $MoS_2$ junction based on homogeneous material of the TMDCs.[25-28]

In this work, we fabricated atomic thin $MoS_2$ junction phototransistors with different $MoS_2$ layers and studied their photoresponse behavior at different source-drain bias and gaseous environment. The difference of band gap in different thickness of few layer $MoS_2$ was utilized to create a built-in electric field. Interestingly, we observed that the short-circuit photocurrent due to the photovoltaic (PV) effect[29] is insensitive to the gaseous environment over the range from a vacuum of $1 \times 10^{-6}$ Torr to ambient condition. This environmental insensitivity can be well ascribed to the unique characteristic of diffusion current that is associated with the carrier density gradient. Conversely, the photocurrent under bias exhibits persistent photoconductivity (PPC) and highly depends on the gaseous environment. The scanning tunneling microscopy and spectroscopy (STM and STS) measurements reveal the energy profile at the MoS$_2$ junction, confirming the presence of the band offset. Moreover, we show that the MoS$_2$ junction phototransistors exhibit gate-voltage tunable open-circuit voltage and short-circuit photocurrent, demonstrating the capability for regulating the current-voltage characteristics of the MoS$_2$ junction via electric field effect.

The details of sample fabrication processes are described in the Supporting Information S1. In brief, MoS$_2$ junctions with different layers were prepared by mechanically exfoliating MoS$_2$ crystals onto SiO$_2$(300 nm)/Si substrates. We then identify MoS$_2$ flakes with different numbers of layers using optical microscopy. The thicknesses of the flakes of MoS$_2$ were characterized using atomic force microscopy and Raman spectroscopy (Supporting Information S2). We then employ a resist-free approach to define the electrical contact to the MoS$_2$ junction to avoid the resist residue which may cause extra carrier scattering. We then deposited Au (50 nm) for the electrical contacts by e-beam evaporation at a base pressure of $1 \times 10^{-7}$ Torr.

**Results and discussion**

We studied the photoresponse behavior of the MoS$_2$ junction devices to investigate the PV effect due to the built-in electric field. Figures 1a and 1b show a schematic of the device structure of MoS$_2$ 1L-3L junction with Au electrodes and an optical microscopy image of a typical MoS$_2$ junction device (sample A), respectively. For sample A, the total channel length is 5 μm, and the thin and thick regions of the MoS$_2$ junctions are 0.7 and 1.9 nm, respectively. We performed the photocurrent measurement with a photoexcitation (532 nm, 20 kW/cm$^2$) focused at the MoS$_2$ junction interface at $V_G = 60$ V. Figure 1c compares the photocurrent at $V_{SD} = 0$ mV and $V_{SD} = 50$ mV, in which we readily found drastically different photoresponse between zero and small $V_{SD}$. For $V_{SD} = 50$ mV, the photocurrent exhibits a PPC which is the photocurrent that persists after the photoexcitation is terminated. For a disordered system, it is common to use the stretched exponential decay to describe the PPC relaxation.[30-32] Here, we can analyze the photocurrent relaxation by a single stretched exponential decay[33]

$$I(t) = I_0 exp[-(t/\tau)^\beta] \qquad (1)$$

where $\tau$ is the decay time, and $\beta$ is the exponent ($0 < \beta < 1$), yielding $\tau = 90$ s for $V_{SD} = 50$ mV. The decay time is comparable to the previous study,[30] suggesting that the observed PPC effect in the MoS$_2$ sample is originated from random localized potential fluctuation.[30-32] In contrast, the short-circuit photocurrent ($I_{SC}$) at zero bias is simply due to the PV effect driven by the built-in electric field, which exhibits a fast switching behavior and returns to the dark current level rapidly after illumination is terminated. The photocurrent relaxation can be fitted by a normal exponential decay ($\beta = 1$ in Eqn. 1).[34-36] The value of $\tau$ is extracted to be approximately 60 ms, which are smaller than that under bias ($V_{SD} = 50$ mV) by 3 orders of magnitude (Supporting information 3).

The drastic difference between photoresponse with and without applying $V_{SD}$ is intriguing and can be attributed to the difference in the carrier transport mechanism.[37] Here, we mainly focus on the slow-changing photocurrent that distinguishes the two scenarios. The current in the channel can be expressed as $J_n = J_n(drift) + J_n(diffusion) = e\mu_n nE - eD_n\, dn/dx$. The carrier density is varied temporally due to the trapping and de-trapping process in the illumination and dark condition, respectively. Because the photoconductivity (PC) is determined by the drift current, which is proportional to the carrier density, the PC is greatly affected by the presence of the charge traps. In contrast, when no bias voltage is applied, $I_{SC}$ is induced by the built-in electric field at the junction and then driven by the diffusion current in the channel, which is proportional to the gradient of the carrier density $dn/dx$. Because $dn/dx$ is related to the gradient of quasi-Fermi level $dF_n/dx$ but not the carrier density, it is much less sensitive to the trapping/de-trapping processes. Therefore, $I_{SC}$ exhibits a much faster photoresponse and the PPC effect can be greatly suppressed when the photocurrent is dominated by the short-circuit current.

Interestingly, the short-circuit photocurrent is virtually insensitive to the variation of the gaseous conditions over a range from a vacuum of $1 \times 10^{-6}$ Torr to ambient condition. Here, we present this unique characteristic by showing the $V_{SD}$ dependent photocurrent ($I_{ph}$) in different gaseous environment. Figure 2a compares $I_{ph}$ of a 1L-3L MoS$_2$ junction (sample B) at $V_{SD} = 0$ mV among vacuum, N$_2$ (1 atm), and ambient condition, revealing that the magnitude of the $I_{SC}$ ($\Delta I_{SC}$) is insensitive to the environment. We note that in all 5 1L-3L samples that we measured, the photoresponse behavior is comparable and therefore the data shown here is representative. Under very different pressure and chemical substances, $\Delta I_{SC}$ exhibits essentially the same value of approximately 140 pA. Moreover, the decay time (τ) values for these different conditions are also similar, as described in detail in supporting information S3. In contrast, the photocurrent of sample B under bias critically depends on the environmental conditions. Figure 2b shows $I_{ph}$ at $V_{SD} = 5$ mV under vacuum, N$_2$, and ambient conditions, indicating the changes in photocurrent ($\Delta I_{PC}$) for these three conditions are 263, 133, and 41 pA, respectively. Here we note that while $\Delta I_{SC}$ is only governed by the built-in electric field, $\Delta I_{PC}$ is referred to PC that is determined by both built-in field and external bias. Moreover, the sample exhibits larger τ under vacuum (90 sec) as compared to the τ under N$_2$ (16 sec) and ambient (8 sec) conditions. This dependence of $I_{ph}$ on the chemical environment is typical for the 2D-material devices. The extracted parameters of mobility, threshold voltage, $\Delta I_{SC}$, and $\Delta I_{PC}$ are listed in Table 1.

As previously discussed, $I_{SC}$ is driven by the diffusion current that is related to the gradient of carrier density. This gradient is associated with the condition of illumination, but is negligibly affected by the trapping process and the presence of adsorbents. Consequently, both the magnitude and the response time of $I_{SC}$ are very insensitive to the gaseous environment, despite the pressure and the chemical composition being very different. We note that this insensitivity of the photocurrent in MoS$_2$ junctions is very uncommon in 2D-material-based devices, considering the ultrahigh surface-to-volume ratio that leads to a large area being exposed to absorbents. Conversely, when bias voltage is applied, the carriers could be trapped or released during illumination and dark conditions, leading to the charging and the PPC effect, as seen in Figure 2b.

|  | Vacuum | N$_2$ | Ambient |
| --- | --- | --- | --- |
| mobility | $0.5\ cm^2/Vs$ | $0.14\ cm^2/Vs$ | $0.09\ cm^2/Vs$ |
| threshold voltage | 5.5 V | 20 V | 25.5 V |
| $\Delta I_{SC}$ V$_{SD}$ = 0 mV | −136 pA | −143 pA | −149 pA |
| $\Delta I_{PC}$ V$_{SD}$ = 5 mV | 263 pA | 133 pA | 41 pA |

**Table 1. The characteristics of the MoS$_2$ junction (sample B) under different gaseous conditions.**

To further investigate the observed PV effect due to the built-in electric field, we present the $V_{SD}$ dependence of the photocurrent. Figure 3a shows a schematic diagram of energy band alignment and photoinduced carrier dynamics in the MoS$_2$ heterojunction. Type-I band alignment is implied based on the STS results shown in Figures 4d and 4e. At zero bias, the photocurrent is mainly driven by the built-in field under illumination (Supporting Information S4), whereas both the built-in and the external field co-exist when bias voltage is applied. It is noted that only the electron conduction is considered here because all the MoS$_2$ junction samples exhibit n-type semiconducting behavior (Supporting Information S1). We considered the photocurrent under two polarity of bias (the detailed calculation of $\Delta I_{SC}$ and $\Delta I_{PC}$ are described in Supporting Information S5). The fast photoresponse of the photocurrent is negative, regardless of the polarity of bias, indicating that $I_{SC}$ is induced by the built-in electric field. We then compared the photocurrent of sample B corresponding to the PV and the charging/discharging effect as a function of $V_{SD}$ in small bias regime, as shown Figure 3b. $\Delta I_{SC}$ is found to be relatively independent of $V_{SD}$; this response is ascribed to the cancellation of the external field. In contrast, $\Delta I_{PC}$, and thus the charging/discharging effect, increases linearly with $V_{SD}$; this response may be due to the phenomenon that the carriers are driven to the local potential minimum more efficiently with increasing bias.

Figure 4a shows a schematic of the experimental setup for the STM measurement on the MoS$_2$ junction transistors. Before MoS$_2$ exfoliation, we deposited a thin TiO$_x$ film of 5 nm,[38] which served as electron transfer dopant on MoS$_2$ flakes,[39] resulting in the reduction of MoS$_2$ sheet resistance. The deposition of a TiO$_x$ layer is therefore crucial for preventing of the STM tip from crashing during the measurement. Figure 4b shows a topography image of the MoS$_2$ junction device (sample C) at the junction, revealing the step with two flat terraces. The apparent step height is 2.6 nm, corresponding to 4 layers of MoS$_2$. Because the thinner MoS$_2$ is characterized by AFM as 4 layers, the studied MoS$_2$ sample is therefore a 4- to 8-layer junction. Figure 4c shows a high magnification topography image of MoS$_2$ that reveals crystalline structure, indicating a pristine MoS$_2$ surface.

To investigate the band alignment at the MoS$_2$ junction, the STS technique was utilized to measure the normalized $dI/dV$ as a function of bias, which corresponds to the local density of state (LDOS). Figure 4d shows the bias dependence of $dI/dV$ curve for the two MoS$_2$ terraces with different thicknesses. The onsets of the normalized $dI/dV$ curves at positive and negative bias correspond to the conduction band edge ($E_C$) and the valance band edge ($E_V$), respectively (Supporting information S6). From the STS measurement, we deduce the following: the $E_C$ and $E_V$ in 8-layer MoS$_2$ are 0.29 eV and -1.12 eV, respectively; the $E_C$ and $E_V$ in 4-layer MoS$_2$ are 0.44 eV and -1.22 eV, respectively. By spatially mapping the normalized $dI/dV$ curves, we plot the band alignment across the 4- to 8-layer junction, as shown in Figure 4e. The deduced energy profile suggests that the band alignment of the MoS$_2$ junction is type I, as depicted in Figure 3a.

Finally, we present the current-voltage characteristics at different $V_G$ to analyze the field effect of the photocurrent in the MoS₂ junction devices. Figure 5a compares the $I_{SD} - V_{SD}$ curves of sample B in dark and under illumination with a 532 nm laser and 20 kW/cm² at $V_G = 60\ V$. The $I_{SD} - V_{SD}$ curves exhibit a linear behavior, indicating that resistor behavior, rather than the rectifying effect, dominates the transport property of the junctions. The slope of the $I_{SD} - V_{SD}$ curves therefore approximates the conductance ($G$) of the device. Under illumination, $G$ is enhanced by 40 times compared with $G$ in dark; this enhancement is attributed to the generation of the photoinduced carriers. The open-circuit voltage ($V_{OC}$) and $I_{SC}$ can be extracted as 60 $\mu$V and 0.12 nA, respectively. We note that the $I_{SC}$ can be further enhanced by reducing the Schottky barrier height via contact engineering.

Because the Fermi level in thin materials can be effectively tuned by the external electric field, it is intriguing to study the field effect of the current-voltage characteristics. Indeed, we observed a $V_G$ dependence of the $I_{SD} - V_{SD}$ curves under illumination, as shown in Figure 5b, in which $G$ increases with increasing $V_G$. To understand this field effect, we plot $V_{OC}$ and $I_{SC}$ as a function of $V_G$, as shown in Figure 5c. $I_{SC}$ is found to increase with increasing $V_G$, similar to the $I_{SD} - V_G$ curve (see Figure S1 of SI 1). This similarity is reasonable because $I_{SC}$ depends on the collection probability of the photoinduced carriers, which is correlated to the diffusion current and $G$. Moreover, as $V_G$ increases, the contact resistance may decrease due to Schottky barrier thinning,[40] leading to higher $I_{SC}$. Conversely, the response of $V_{OC}$ decreasing with increasing $V_G$ can be attributed to the reduction of built-in electric field in the MoS₂ junction. Because monolayer MoS₂ is subjected to a stronger field effect compared with the few-layer MoS₂ due to the different density of states, the rising of Fermi level in monolayer MoS₂ is greater than that in few-layer MoS₂ as $V_G$ increases, resulting in the reduction of the built-in electric field (see Figure 3a) and thus $V_{OC}$.

We further present the excitation power dependence of $I_{SC}$ at zero bias voltage (Figure 5d) to examine the mechanism of the photocurrent generation.[41] We observe that $I_{SC}$ follows a power law $I_{SC} \propto P^{\alpha}$, and the exponent α can be been extracted as 0.89 and 0.82 for $V_G = 0$ and 60 V, respectively. For the PC and the PV process, the photoinduced carrier density is directly proportional to the rate of absorbed photons; therefore, $\alpha = 1$.[42] However, because the PC is excluded here ($V_{SD} = 0$), the value of α therefore suggests that the PV is the dominant mechanism in the measured photocurrent. The deviation of the extracted α from unity may be attributed to electron-hole recombination at the $MoS_2$ junction and/or nonradiative recombination centers.[43,44]

In conclusion, we demonstrated a unique environment-insensitive and gate-controllable short-circuit photocurrent in a $MoS_2$ junction with differences in the number of layer. The environmental insensitivity of the short-circuit photocurrent can be attributed to the characteristic of the diffusion current. Conversely, the photocurrent with bias exhibits the typical PPC that greatly depends on the amount of the extrinsic adsorbents. The STM/STS measurement confirms the quality of the $MoS_2$ junction samples and suggests the type-I band alignment of the junction. In addition to the effect of source-drain bias, the $MoS_2$ junction devices exhibit strong back-gate voltage dependence, indicating the feasibility to control the photocurrent via field effect. The environment-insensitive photocurrent therefore shows an alternative method to design the device structure for TMDC-based electronic and optoelectronic applications.

**Methods**

**Sample preparation**. The sample with the 1L-3L $MoS_2$ junction was produced via mechanical exfoliation of $MoS_2$ layers from the bulk $MoS_2$ (SPI supplies) onto $SiO_2$ (300 nm)/Si substrates. Next, a resist-free technique with a shadow mask (TEM grids) was utilized to deposit electrical contacts. The advantage of the resist-free technique is the lack of resist residue on the $MoS_2$ surface resulting from the device fabrication process. We deposited Au (50 nm) as the electrical contacts using an electron-beam evaporator at a base pressure of $1.0 \times 10^{-7}$ Torr. All of our $MoS_2$ junction devices were measured in a cryostat (Janis Research Company, ST-500) under vacuum condition of $1.0 \times 10^{-6}$ Torr. We performed DC electrical measurement using a Keithley 237 sourcemeter and applied the back-gate voltage using a Keithley 2400 sourcemeter. We employed solid-state CW laser (Nd:YAG, 532 nm) as the light source in the Raman spectroscopy and photoresponse measurements. The incident light beam was focused by an objective (100×, NA 0.6) with a spot size of ~ 0.9 μm.

## References

1   Geim, A. K. & Novoselov, K. S. The rise of graphene. *Nature materials* **6**, 183-191 (2007).
2   Schedin, F. *et al.* Detection of individual gas molecules adsorbed on graphene. *Nature materials* **6**, 652-655 (2007).
3   Varghese, S. S., Varghese, S. H., Swaminathan, S., Singh, K. K. & Mittal, V. Two-Dimensional Materials for Sensing: Graphene and Beyond. *Electronics* **4**, 651-687 (2015).
4   Tongay, S. *et al.* Broad-range modulation of light emission in two-dimensional semiconductors by molecular physisorption gating. *Nano letters* **13**, 2831-2836 (2013).
5   He, R. *et al.* Large physisorption strain in chemical vapor deposition of graphene on copper substrates. *Nano letters* **12**, 2408-2413 (2012).
6   Perkins, F. K. *et al.* Chemical vapor sensing with monolayer MoS2. *Nano letters* **13**, 668-673 (2013).
7   Ando, T. Screening effect and impurity scattering in monolayer graphene. *Journal of the Physical Society of Japan* **75**, 074716 (2006).
8   Hwang, E., Adam, S. & Sarma, S. D. Carrier transport in two-dimensional graphene layers. *Physical Review Letters* **98**, 186806 (2007).
9   Bolotin, K. I. *et al.* Ultrahigh electron mobility in suspended graphene. *Solid State Communications* **146**, 351-355 (2008).
10  Ma, N. & Jena, D. Charge scattering and mobility in atomically thin semiconductors. *Physical Review X* **4**, 011043 (2014).
11  Furchi, M. M., Polyushkin, D. K., Pospischil, A. & Mueller, T. Mechanisms of Photoconductivity in Atomically Thin MoS2. *Nano Lett.* **14**, 6165-6170, doi:10.1021/nl502339q (2014).
12  Chen, J.-H. *et al.* Charged-impurity scattering in graphene. *Nature Physics* **4**, 377-381 (2008).
13  Tan, Y.-W. *et al.* Measurement of scattering rate and minimum conductivity in graphene. *Physical review letters* **99**, 246803 (2007).
14  Chen, J.-H., Jang, C., Xiao, S., Ishigami, M. & Fuhrer, M. S. Intrinsic and extrinsic performance limits of graphene devices on SiO2. *Nature nanotechnology* **3**, 206-209 (2008).
15  Jena, D. & Konar, A. Enhancement of carrier mobility in semiconductor nanostructures by dielectric engineering. *Physical review letters* **98**, 136805 (2007).
16  Kretinin, A. *et al.* Electronic properties of graphene encapsulated with different two-dimensional atomic crystals. *Nano letters* **14**, 3270-3276 (2014).

17  Ho, P.-H. *et al.* Self-encapsulated doping of n-type graphene transistors with extended air stability. *ACS nano* **6**, 6215-6221 (2012).
18  Lee, G.-H. *et al.* Highly Stable, Dual-Gated MoS2 Transistors Encapsulated by Hexagonal Boron Nitride with Gate-Controllable Contact, Resistance, and Threshold Voltage. *ACS nano* **9**, 7019-7026 (2015).
19  Li, L. *et al.* Quantum Hall effect in black phosphorus two-dimensional electron system. *Nature Nanotechnology* (2016).
20  Kuc, A., Zibouche, N. & Heine, T. Influence of quantum confinement on the electronic structure of the transition metal sulfide T S 2. *Physical Review B* **83**, 245213 (2011).
21  Yun, W. S., Han, S., Hong, S. C., Kim, I. G. & Lee, J. Thickness and strain effects on electronic structures of transition metal dichalcogenides: 2H-M X 2 semiconductors (M= Mo, W; X= S, Se, Te). *Physical Review B* **85**, 033305 (2012).
22  Mak, K. F., Lee, C., Hone, J., Shan, J. & Heinz, T. F. Atomically thin MoS 2: a new direct-gap semiconductor. *Physical Review Letters* **105**, 136805 (2010).
23  Splendiani, A. *et al.* Emerging photoluminescence in monolayer MoS2. *Nano letters* **10**, 1271-1275 (2010).
24  Cheiwchanchamnangij, T. & Lambrecht, W. R. Quasiparticle band structure calculation of monolayer, bilayer, and bulk MoS 2. *Physical Review B* **85**, 205302 (2012).
25  Howell, S. L. *et al.* Investigation of Band-Offsets at Monolayer–Multilayer MoS2 Junctions by Scanning Photocurrent Microscopy. *Nano letters* **15**, 2278-2284 (2015).
26  Tosun, M. *et al.* MoS2 Heterojunctions by Thickness Modulation. *Scientific reports* **5** (2015).
27  Baugher, B. W. H., Churchill, H. O. H., Yang, Y. F. & Jarillo-Herrero, P. Optoelectronic devices based on electrically tunable p-n diodes in a monolayer dichalcogenide. *Nat Nanotechnol* **9**, 262-267, doi:10.1038/Nnano.2014.25 (2014).
28  Pospischil, A., Furchi, M. M. & Mueller, T. Solar-energy conversion and light emission in an atomic monolayer p-n diode. *Nat Nanotechnol* **9**, 257-261, doi:10.1038/Nnano.2014.14 (2014).
29  Fontana, M. *et al.* Electron-hole transport and photovoltaic effect in gated $MoS_2$ Schottky junctions. *Sci Rep-Uk* **3**, doi:Artn 1634

Doi 10.1038/Srep01634 (2013).
30  Wu, Y. C. *et al.* Extrinsic Origin of Persistent Photoconductivity in Monolayer MoS2 Field Effect Transistors. *Sci Rep* **5**, 11472, doi:10.1038/srep11472 (2015).
31  Jiang, H. X. & Lin, J. Y. Percolation transition of persistent photoconductivity in II-VI mixed crystals. *Phys Rev Lett* **64**, 2547-2550, doi:DOI 10.1103/PhysRevLett.64.2547 (1990).
32  Dissanayake, A. S., Huang, S. X., Jiang, H. X. & Lin, J. Y. Charge Storage and Persistent Photoconductivity in a Cds0.5se0.5 Semiconductor Alloy. *Phys Rev B* **44**, 13343-13348, doi:DOI 10.1103/PhysRevB.44.13343 (1991).
33  Palmer, R. G., Stein, D. L., Abrahams, E. & Anderson, P. W. Models of Hierarchically Constrained Dynamics for Glassy Relaxation. *Phys. Rev. Lett.* **53**, 958-961 (1984).
34  Yin, Z. Y. *et al.* Single-Layer $MoS_2$ Phototransistors. *ACS Nano* **6**, 74-80, doi:Doi 10.1021/Nn2024557 (2012).
35  Lee, H. S. *et al.* $MoS_2$ Nanosheet Phototransistors with Thickness-Modulated Optical Energy Gap. *Nano Lett.* **12**, 3695-3700, doi:Doi 10.1021/Nl301485q (2012).
36  Chen, S. Y. *et al.* Biologically inspired graphene-chlorophyll phototransistors with high gain. *Carbon* **63**, 23-29, doi:10.1016/j.carbon.2013.06.031 (2013).
37  Sze, S. M. & Ng, K. K. *Physics of semiconductor devices*. (John wiley & sons, 2006).
38  Ho, P. H. *et al.* Precisely Controlled Ultrastrong Photoinduced Doping at Graphene–Heterostructures Assisted by Trap-State-Mediated Charge Transfer. *Advanced materials* **27**, 7809-7815 (2015).

39    Rai, A. *et al.* Air stable doping and intrinsic mobility enhancement in monolayer molybdenum disulfide by amorphous titanium suboxide encapsulation. *Nano letters* **15**, 4329-4336 (2015).
40    Li, S.-L. *et al.* Thickness Scaling Effect on Interfacial Barrier and Electrical Contact to Two-Dimensional MoS2 Layers. *ACS Nano*, doi:10.1021/nn506138y (2014).
41    Zhang, Y. *et al.* Photothermoelectric and photovoltaic effects both present in MoS2. *Scientific reports* **5**, 7938, doi:10.1038/srep07938 (2015).
42    Patil, V., Capone, A., Strauf, S. & Yang, E.-H. Improved photoresponse with enhanced photoelectric contribution in fully suspended graphene photodetectors. *Scientific reports* **3** (2013).
43    Zhang, W. *et al.* High-gain phototransistors based on a CVD MoS(2) monolayer. *Advanced materials* **25**, 3456-3461, doi:10.1002/adma.201301244 (2013).
44    Cho, K. *et al.* Gate-bias stress-dependent photoconductive characteristics of multi-layer MoS2 field-effect transistors. *Nanotechnology* **25**, 155201, doi:10.1088/0957-4484/25/15/155201 (2014).

ACKNOWLEDGMENT

This work was supported by the Ministry of Science and Technology, Taiwan under contract numbers MOST 103-2112-M-001-020-MY3.

**Author Contributions Statement**

W.H.W. supervised the project. W.H.W., F.Y.S., and Y.C.W. designed the experiments. Y.S.S., M.C.S., and Y.P.C. performed the STM and STS measurement and analysis. P.H.H. and C.W.C. provided the OTS-functionalized substrates. F.Y.S., Y.C.W., and T.S.W. prepared the samples and carried out the photoresponse and transport measurements. F.Y.S., Y.C.W., and T.S.W. analyzed the data. W.H.W, F.Y.S., C.W.C., Y.F.C. and Y.P.C. wrote the paper. All authors discussed the results and contributed to the refinement of the paper.

**Additional information**

**Competing financial interests:** The authors declare no competing financial interests.

**Figure Legends**

**Figure 1. The structure and the photoresponse behaviors of the MoS₂ junctions.** (a) A schematic diagram of a 1L-3L MoS₂ junction transistor with the excitation beam focused on the MoS₂ junction. (b) Optical image of a MoS₂ junction transistor. The edge of the MoS₂ junction is outlined by red lines. (c) Time-resolved photoresponse behaviors of the MoS₂ junction transistor (sample A) under $V_{SD} = 50$ mV (red curve) and $V_{SD} = 0$ mV (blue curve) at $V_G = 60$ V.

**Figure 2. Time-resolved photoresponse of the MoS₂ junctions.** The photocurrent behavior of the MoS₂ junction transistor (sample B) under different gaseous conditions at (a) $V_{SD} = 0$ mV and (b) $V_{SD} = 5$ mV. The short-circuit photoresponse is virtually insensitive to the variation of the gaseous conditions.

**Figure 3. V_SD-dependent photoresponse of the MoS₂ junctions.** (a) A schematic of the band structure of the MoS₂ junction transistor and photoinduced carrier transfer at $V_{SD} = 0$ V and $V_{SD} \neq 0$ V. (b) The photocurrent measurement of sample B corresponding to the photovoltaic effect (black squares) and the photoconductivity effect (red squares) as a function of $V_{SD}$ at $V_G = 60$ V. The excitation power is 200 μW.

**Figure 4. The STM/STS measurement of the MoS₂ junctions.** (a) A schematic of the MoS₂ junction structure with different layer in the STM/STS measurement. (b) Top: STM topography image of the MoS₂ junction (sample C). Bottom: a cross-sectional topographic profile of the MoS₂ junction. (c) An STM image of sample C with atomic-scale resolution, which indicates a pristine MoS₂ surface. (d) Normalized $dI/dV$ curves of 4 layer (green curve) and 8 layer (red curve) MoS₂. The profiles are offset for clarity. (e) Band alignment across the MoS$_2$ 4- to 8-layer junction. Type-I band alignment at the MoS$_2$ junction was implied.

**Figure 5. Field-effect-controlled short-circuit photocurrent of the MoS$_2$ junction transistors.** (a) Output curves of the MoS$_2$ junction device (sample B) in dark (black curve) and under 532 nm excitation focused on the MoS$_2$ junction (red curve) at $V_G = 60$ V. (b) $V_G$ dependence of the output curves under laser illumination (P = 200 μW) at the MoS$_2$ junction. (c) Analysis of $V_G$ dependence of $V_{OC}$ and $I_{SC}$ extracted from the output curves. (d) Excitation power dependence of $I_{SC}$ at zero bias at $V_G = 0$ V (black squares) and $V_G = 60$ V (red circles). The dashed lines are fitting curves of power law $I_{SC} \propto P^\alpha$.

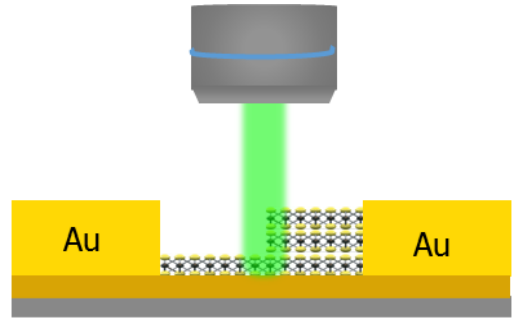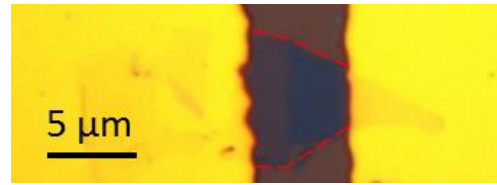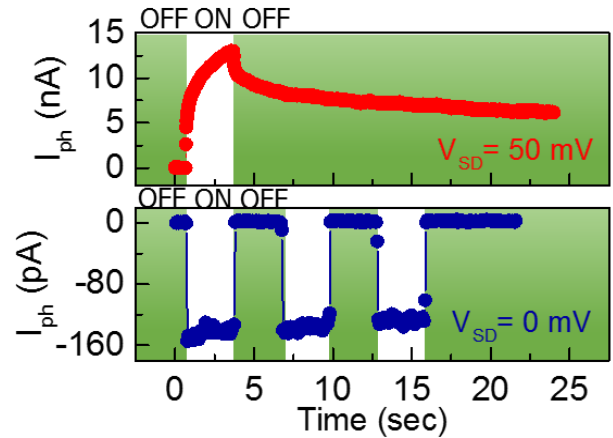

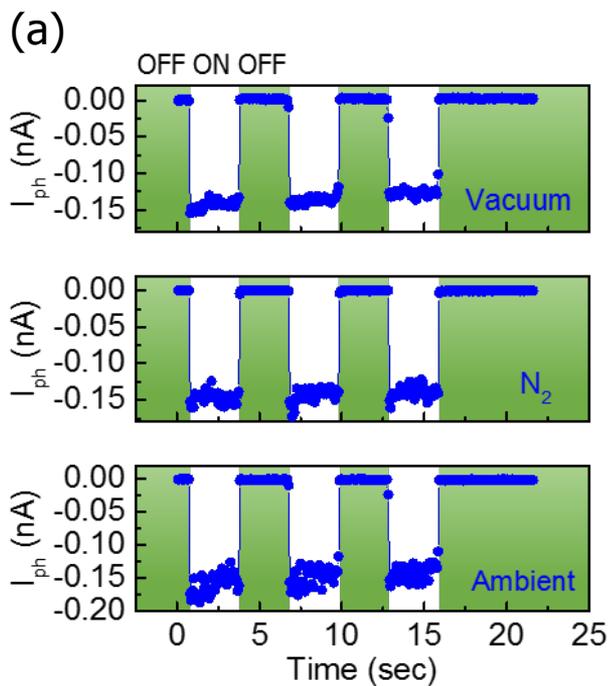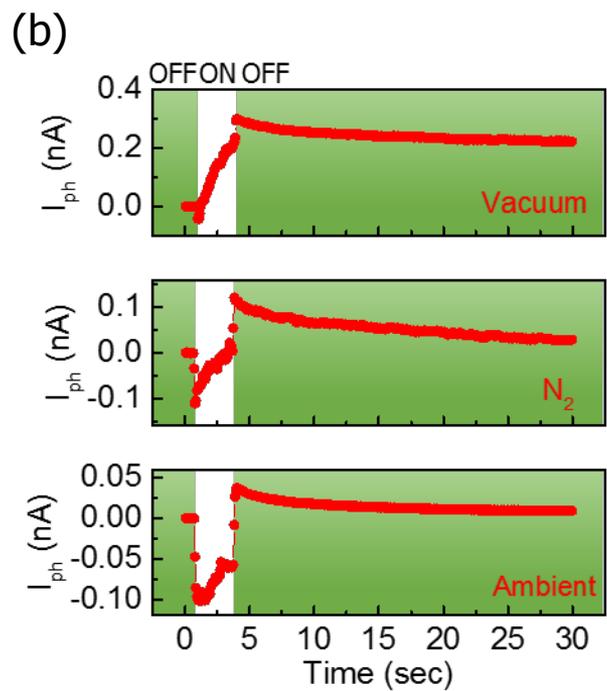

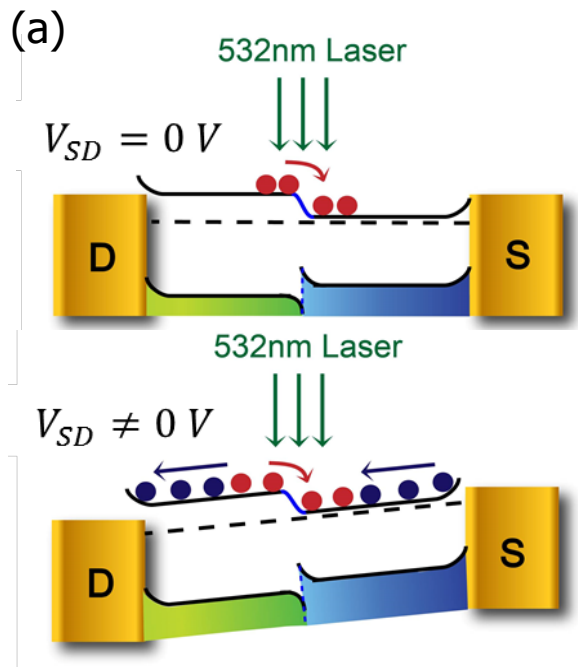 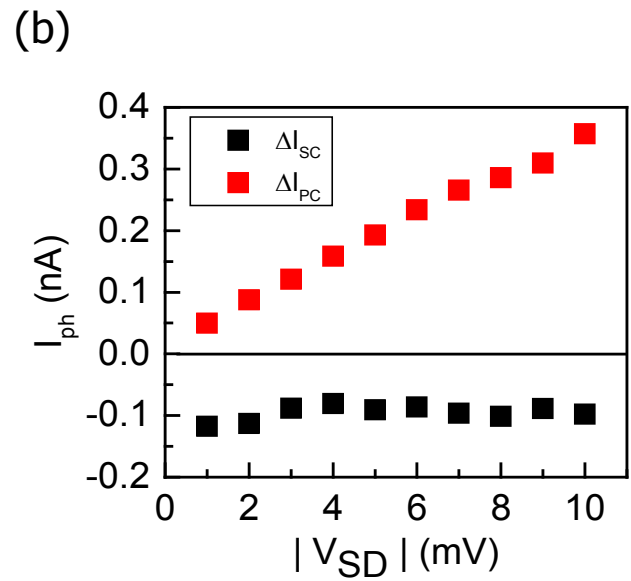

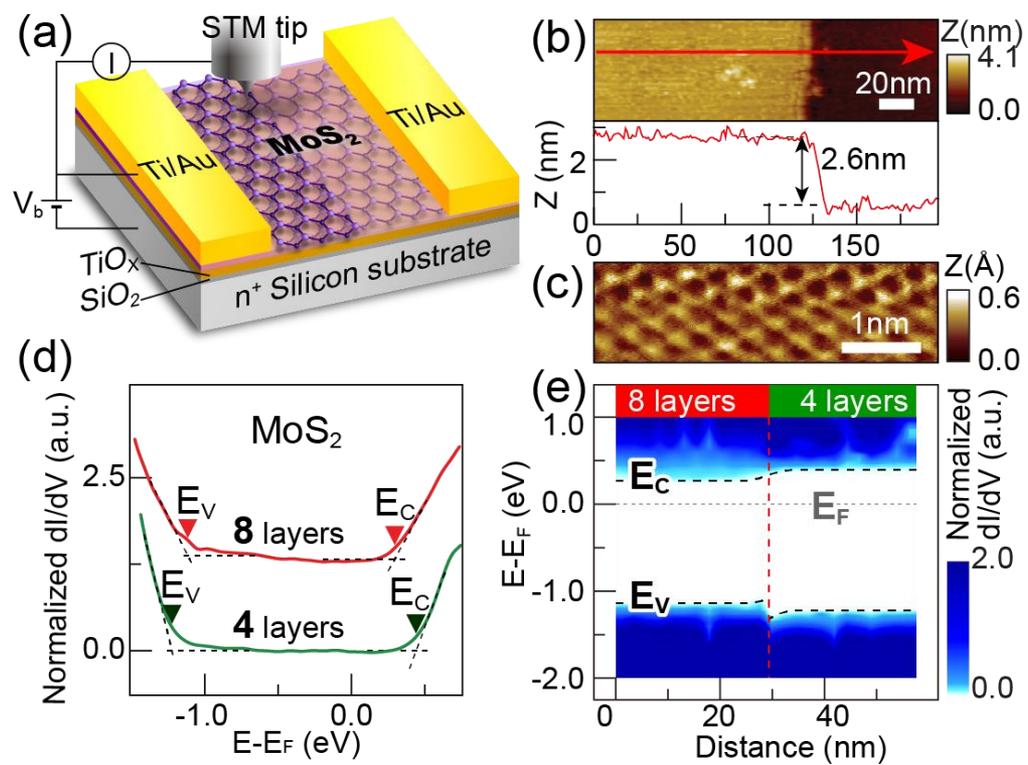

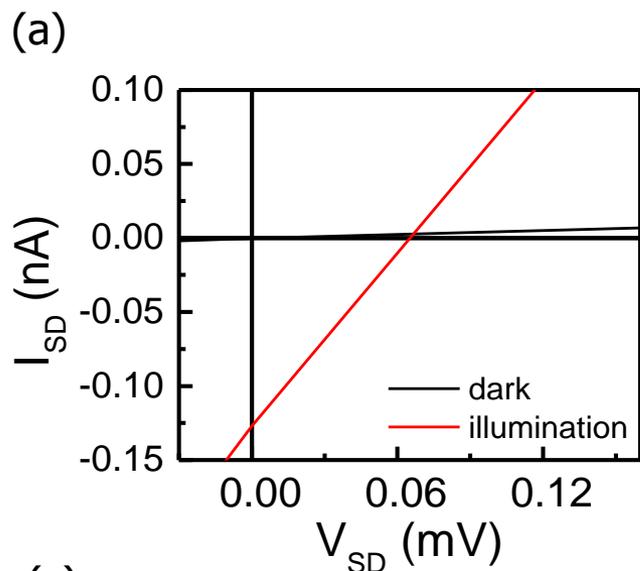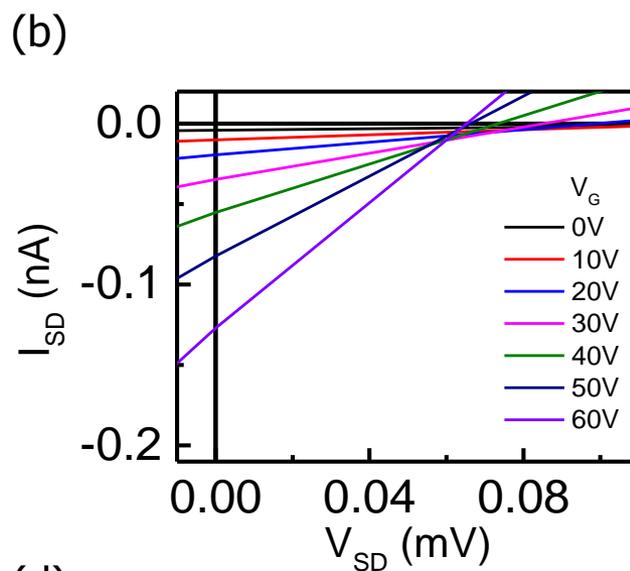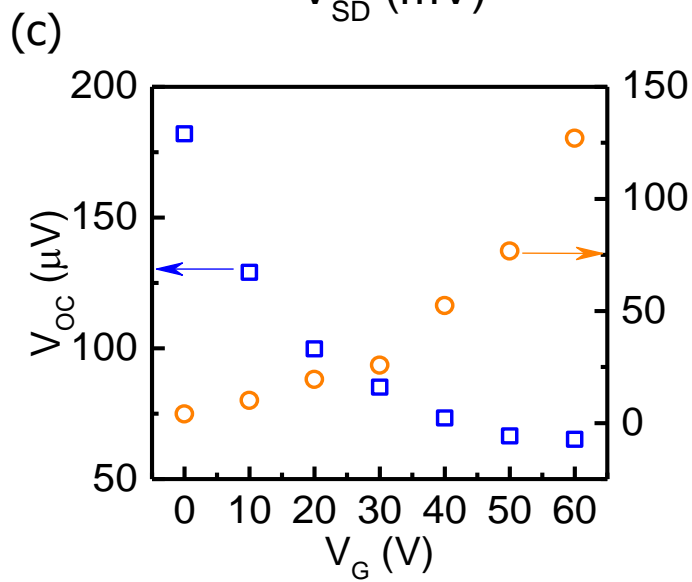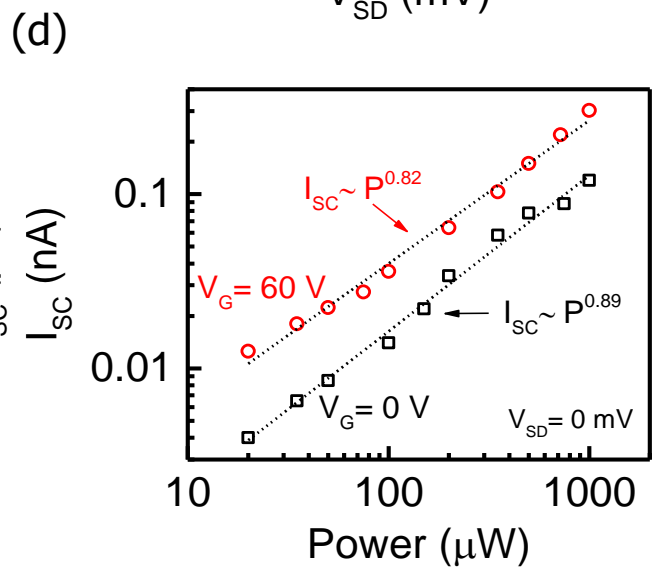

Supporting Information

# Environment-insensitive and gate-controllable photocurrent enabled by bandgap engineering of MoS$_2$ junctions

Fu-Yu Shih[1,2], Yueh-Chun Wu[2,§], Yi-Siang Shih[1], Ming-Chiuan Shih[3], Tsuei-Shin Wu[1,2], Po-Hsun Ho[4], Chun-Wei Chen[4], Yang-Fang Chen[1], Ya-Ping Chiu[1,5], and Wei-Hua Wang[2*]

[1]Department of Physics, National Taiwan University, Taipei 106, Taiwan

[2]Institute of Atomic and Molecular Sciences, Academia Sinica, Taipei 106, Taiwan

[3]Department of Physics, National Sun Yat-sen University, Kaohsiung, Taiwan

[4]Department of Materials Science and Engineering, National Taiwan University, Taipei 106, Taiwan

[5]Institute of Physics, Academia Sinica, Taipei 115, Taiwan

[§]Current address: Department of Physics, University of Massachusetts, Amherst, Massachusetts 01003, United States

[*]Corresponding Author. (W.-H. Wang) Tel: +886-2-2366-8208, Fax: +886-2-2362-0200;

E-mail: wwang@sinica.edu.tw

**S1. Device fabrication and photocurrent measurement**

The sample with the 1L-3L MoS$_2$ junction was produced via mechanical exfoliation of MoS$_2$ layers from the bulk MoS$_2$ (SPI supplies) onto SiO$_2$ (300 nm)/Si substrates. Next, a resist-free technique with a shadow mask (TEM grids) was utilized to deposit electrical contacts. The advantage of the resist-free technique is the lack of resist residue on the MoS$_2$ surface resulting from the device fabrication process. We deposited Au (50 nm) as the electrical contacts using an electron-beam evaporator at a base pressure of $1.0 \times 10^{-7}$ Torr. All of our MoS$_2$ junction devices were measured in a cryostat (Janis Research Company, ST-500) under vacuum condition of $1.0 \times 10^{-6}$ Torr. We performed DC electrical measurement using a Keithley 237 sourcemeter and applied the back-gate voltage using a Keithley 2400 sourcemeter. We employed solid-state CW laser (Nd:YAG, 532 nm) as the light source in the Raman spectroscopy and photoresponse measurements. The incident light beam was focused by an objective (100×, NA 0.6) with a spot size of ~ 0.9 µm.

Figure S1 compares the $I_{SD} - V_G$ curves of a 1L-3L MoS$_2$ junction device (sample B of the main text) under vacuum, N$_2$, and ambient conditions. The MoS$_2$ junction device exhibits typical n-type semiconducting behavior. The sample exhibits higher mobility under vacuum (0.5 cm$^2$/Vs) compared to the mobility under N$_2$ (0.14 cm$^2$/Vs) and ambient (0.09 cm$^2$/Vs) conditions. This difference can be understood because the carrier scattering is higher when additional adsorbents are introduced in the environment with abundant molecules [1]. The on/off ratio is approximately 10$^4$. The threshold for N$_2$ and ambient conditions is higher than that under vacuum, suggesting a p-type doping effect is introduced during the adsorption [2, 3].

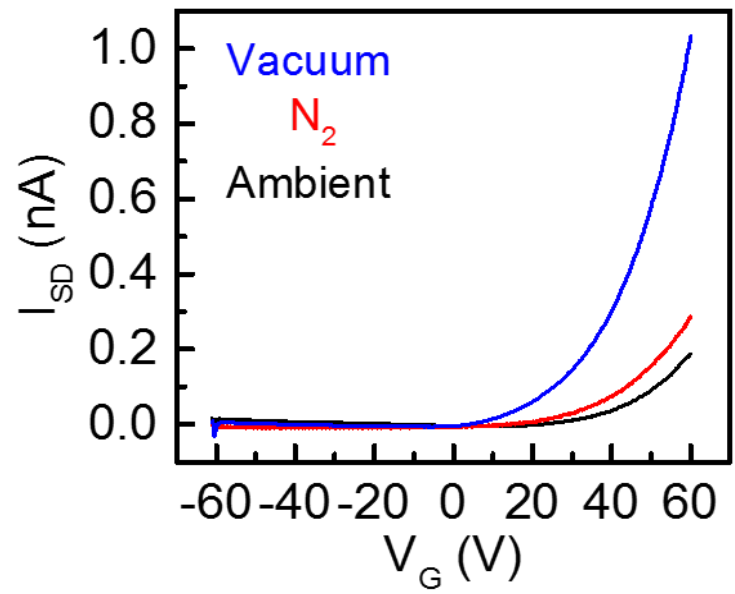

Figure S1. The transfer curves of 1L-3L MoS$_2$ junction device under vacuum, N$_2$, and ambient conditions.

# S2 Identification of the number of MoS₂ layers

The layer number of the MoS$_2$ flakes was identified using optical microscopy, Raman spectroscopy, and atomic force microscopy (AFM) measurements. Figure S2a presents an optical image of MoS$_2$ flake on 300-nm SiO$_2$/Si substrate involving monolayer and trilayer MoS$_2$ (sample A of the main text). The Raman spectra (Figure S2b) reveals two characteristic peaks 388.1 (386.6) cm$^{-1}$ and 407 (409.1) cm$^{-1}$ of the monolayer (trilayer) MoS$_2$ flake, which correspond to the $E_{2g}^1$ and $A_{1g}$ resonance modes, respectively. The difference between the two peaks can be used to identify the thickness of the MoS$_2$, particularly for a number of layers less than 3 [4]. For this MoS$_2$ junction sample, the difference is equal to 18.9 cm$^{-1}$ for monolayer MoS$_2$ and 22.5 cm$^{-1}$ for trilayer MoS$_2$, in agreement with previous reports [4]. We performed AFM measurement in a region denoted by a red rectangle shown in Figure S2a. The result is presented in Figure S2c, which shows that the step height of the thinner area of the MoS$_2$ junction was approximately 0.7 nm, which corresponds to one atomic layer. The step height of the thicker area of the MoS$_2$ junction was approximately 1.9 nm, which corresponds to trilayer MoS$_2$.

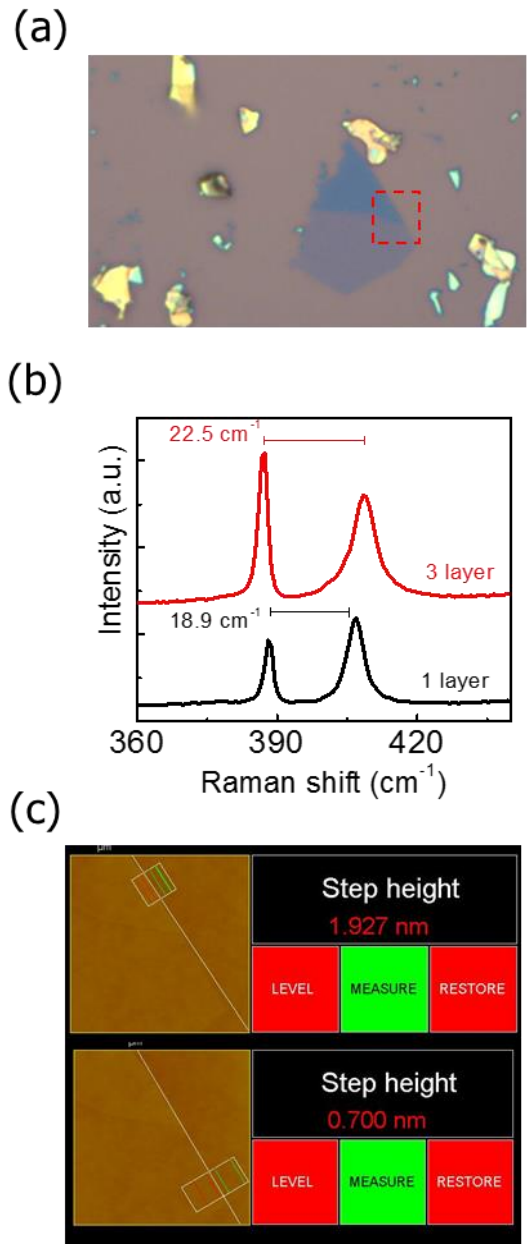

Figure S2. (a) Optical microscope image and (b) Raman spectra and (c) AFM image of 1L-3L MoS$_2$ junction (Sample A).

**S3 Estimation of the decay time under different gaseous conditions**

We measured the time-resolved photocurrent by using the built-in sweep function of the Keithley 237 sourcemeter with the setting-reading cycle of 18 ms. Figure S3a shows the time-resolved photocurrent (black squares) at zero bias voltage under ambient, $N_2$, and vacuum condition, respectively along with the fitting curves (red lines) based on normal exponential decay, $I(t) = I_0 exp-(t/\tau)$. The values of $\tau$ under ambient, $N_2$, and vacuum condition were extracted as 61.8 ms, 63.8 ms, and 62.1 ms, respectively. The value of $\tau$ is found to be approximately the same under different environmental conditions. Figure S3b shows time-resolved photocurrent at $V_{SD} = 5\ mV$ under ambient, $N_2$ and vacuum conditions along with the fitting curves (green lines) based on stretched exponential decay, $I(t) = I_0 exp[-(t/\tau)^\beta]$. The extracted $\tau$ is found to be dependent on the environment; this result can be attributed to the random localized potential fluctuations in $MoS_2$ [5].

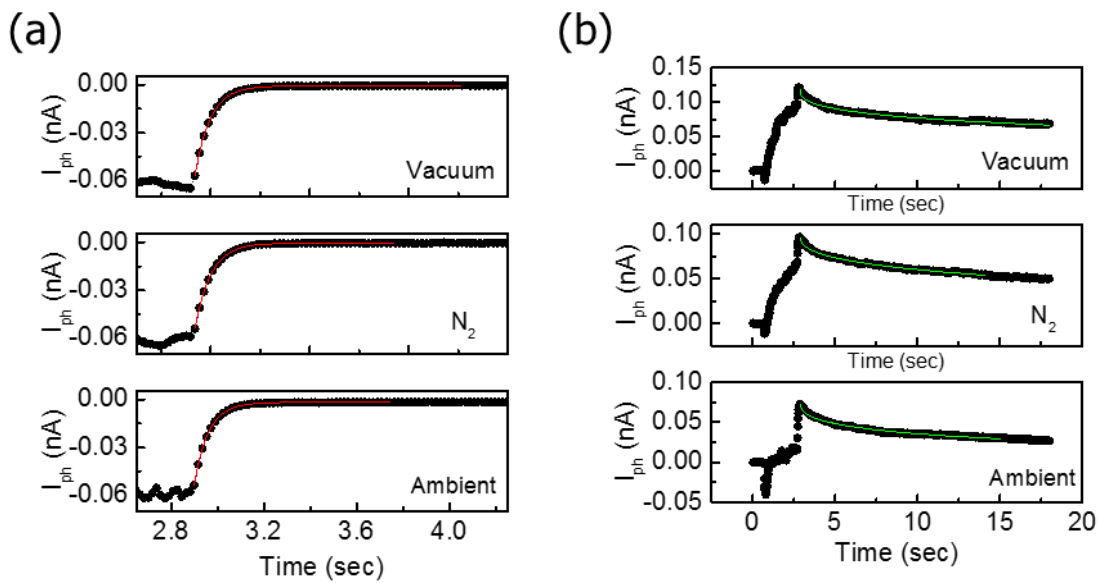

Figure S3. Time-resolved photocurrent measurement at (a) zero bias and (b) $V_{SD} = 5\ mV$ under different gaseous conditions. The red and green curves represent the fitting curves based on normal exponential decay and stretched exponential decay, respectively.

|  | Vacuum | N$_2$ | Ambient |
|---|---|---|---|
| decay time ($\tau$) V$_{SD}$ = 0 mV | 62.1 ms | 63.8 ms | 61.8 ms |
| decay time ($\tau$) V$_{SD}$ = 5 mV | 64 s | 37.4 s | 15.2 s |

Table S1. The decay time of the MoS$_2$ junction (sample B) under different gaseous conditions.

## S4 Zero-bias time-resolved photoresponse in a uniform MoS$_2$ transistor

The observed zero-bias photocurrent in the MoS$_2$ junction device is due to the band offset resulting from the different thicknesses of MoS$_2$. To verify this characteristic, we present the zero-bias photocurrent measurement in a uniform MoS$_2$ device, which shows no photoresponse when illuminated at zero bias (Figure S4). This lack of photoresponse is in contrast to the observed photoresponse measured in the MoS$_2$ junction device, as shown in Figures 1c and 2a in the main text.

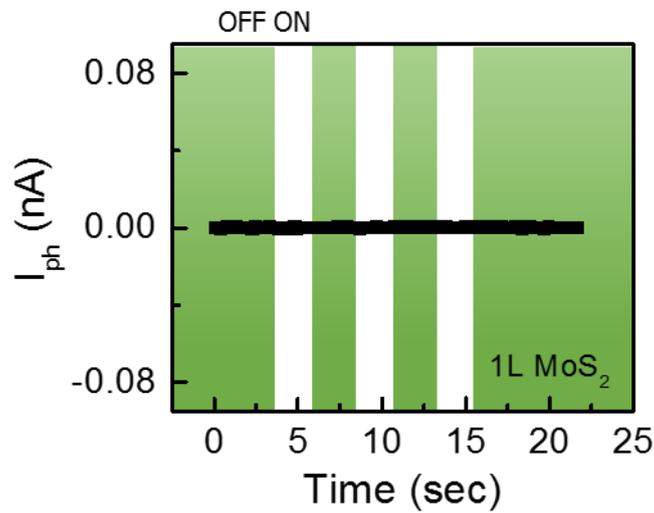

Figure S4. Time-resolved photoresponse of uniform 1L MoS$_2$ at zero-bias voltage.

**S5 Analysis of $V_{SD}$ dependence of the photocurrent due to the PV and the PC effect**

Figures S5a and S5b show the temporal photoresponse of the 1L-3L MoS$_2$ junction device under small bias ($V_{SD} = \pm 2$ mV). The direction of the photocurrent with slow response is found to depend on the direction of the external electric field. In contrast, the direction of the photocurrent with quick response is observed to retain a specific direction, suggesting that the photocurrent with quick response is mainly caused by the junction-induced built-in electric field.

To distinguish the photocurrent corresponding to the PV effect versus the PC effect, we define $I_{SC}$ and $I_{PC}$ as:

$$I_{SC} = \left(I_{V_{SD}=x\ mV} + I_{V_{SD}=-x\ mV}\right)/2 \quad \text{Eqn. S1}$$

$$I_{PC} = \left(\left|I_{PC,V_{SD}=x\ mV}\right| + \left|I_{PC,V_{SD}=-x\ mV}\right|\right)/2 \quad \text{Eqn. S2}$$

where $I_{V_{SD}=x\ mV}$ is the change of photocurrent with quick response at $V_{SD} = x\ mV$, and $I_{PC,V_{SD}=x\ mV}$ is the change of photocurrent with slow response at $V_{SD} = x\ mV$. The photocurrent caused by external bias with quick response is cancelled by summing up the contributions from the two polarities of the bias voltages. Therefore, $I_{SC}$ represents the photocurrent resulting from the built-in electric field. Alternatively, we take the average of the absolute value of the photocurrent with slow response under the two polarities of bias to represent the photocurrent due to the PC effect.

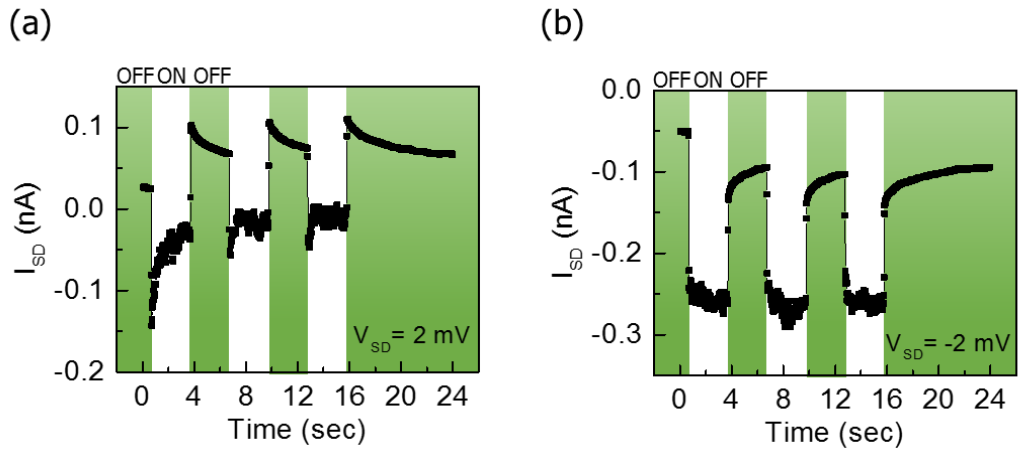

Figure S5. Time-resolved photocurrent of 1L-3L MoS$_2$ junction under small bias ($V_{SD} = \pm 2$ mV).

## S6 STM/STS characterization

In STS measurements, the STM probe tip, vacuum and MoS$_2$ form a metal-insulator-semiconductor (MIS) tunneling junction. When we apply zero sample bias on the sample so that there is no tunneling current between tip and MoS$_2$, the Fermi-level of the tip is equal to the Fermi-level of the MoS$_2$ under thermal equilibrium. Therefore, the zero sample bias can be defined as the Fermi-level of the sample. In contrast to the zero sample bias condition, when we apply a certain sample bias on the sample, the tunneling current flows between the probe tip and MoS$_2$. As a result, the onsets in the normalized $dI/dV$ curves indicate the current onsets from the valance band edge ($E_V$) and the conduction band edge ($E_C$) in STM measurements.

# References

1. Qiu, H., et al., *Electrical characterization of back-gated bi-layer MoS2 field-effect transistors and the effect of ambient on their performances.* Applied Physics Letters, 2012. **100**(12): p. 123104.
2. Tongay, S., et al., *Broad-range modulation of light emission in two-dimensional semiconductors by molecular physisorption gating.* Nano letters, 2013. **13**(6): p. 2831-2836.
3. Yue, Q., et al., *Adsorption of gas molecules on monolayer MoS2 and effect of applied electric field.* Nanoscale research letters, 2013. **8**(1): p. 1-7.
4. Li, H., et al., *From bulk to monolayer MoS2: evolution of Raman scattering.* Advanced Functional Materials, 2012. **22**(7): p. 1385-1390.
5. Wu, Y.-C., et al., *Extrinsic Origin of Persistent Photoconductivity in Monolayer MoS2 Field Effect Transistors.* arXiv preprint arXiv:1505.05234, 2015.